\documentclass[aps,prl,preprint,groupedaddress]{revtex4}
\usepackage{graphics}
\usepackage{graphicx}
\usepackage{amssymb}
\usepackage{color}

\begin{document}

\bibliographystyle{apsrev}

\title{Unexpectedly wide reversible vortex region in $\beta$-pyrochlore RbOs$_{2}$O$_{6}$}

\author{Pierre Legendre, Yanina Fasano, Ivan Maggio-Aprile, {\O}ystein Fischer}

\affiliation{D\'epartement de Physique de la Mati\`ere Condens\'
ee, Universit\'e de Gen\`eve, 24 Quai Ernest-Ansermet, 1211
Geneva, Switzerland}

\author{Zbigniew Bukowski, Sergiy Katrych, Janusz Karpinski}

\affiliation{Laboratory for Solid State Physics, ETH Z\"urich,
8093 Z\"urich, Switzerland}


\date{\today}

\begin{abstract}

We study the extent of the reversible region in the vortex phase
diagram of recently available RbOs$_2$O$_6$ single crystals
\cite{Rogacki08a} by means of bulk magnetization measurements. We
found that the irreversible magnetic response sets in at a field
$H_{\rm irr}(T) \sim 0.3 H_{\rm c2}(T)$ for $0.5 \lesssim T/T_{\rm
c} \lesssim 0.8$, yielding a reversible vortex region that is
 wide in comparison with other low-$T_{\rm c}$
materials. The relevance of thermal fluctuations is limited since
we estimate a Ginzburg number $G_{\rm i} = 5 \times 10^{-7}$.
However, the relevance of quenched disorder is low since the
critical-current density ratio at low fields and temperatures is
of the order of that found in high-T$_{\rm c}$'s. We therefore
conclude that an intrinsically low bulk pinning magnitude favors
the existence of an unexpectedly wide reversible vortex region in
RbOs$_{2}$O$_{6}$.

\end{abstract}

\pacs{74.25.Dw, 74.25.Ha, 74.25.Sv} \keywords{Rb
$\beta$-pyrochlore, magnetic properties, reversible vortex region}

\maketitle

\section{Introduction}

The recent discovery of superconductivity in the
$\beta$-pyrochlore osmate compounds AOs$_{2}$O$_{6}$ (A= K
\cite{Yonezawa04a,Schuck06a}, Rb
\cite{Yonezawa04b,Kazakov04a,Bruhwiler04a}, Cs \cite{Yonezawa04c})
triggered a plethora of studies on the superconducting pairing
mechanism in these compounds. In spite of the active research on
$\beta$-pyrochlores over the last four years, only three works
report on the magnetic and spectroscopic properties of vortex
matter in these compounds \cite{Hiroi06a,Shibauchi07a,Dubois07a}.
These three studies focus on KOs$_{2}$O$_{6}$, the
$\beta$-pyrochlore which presents the highest $T_{\rm c} \sim
9.6$\,K \cite{Hiroi05a}. This compound has a Ginzburg-Landau
parameter $\kappa =70-87$ \cite{Khasanov04a,Bruhwiler06a} and
$H_{\rm c2}(0)=24$ (single-crystals \cite{Bruhwiler06a}) - 33\,T
(polycrystals \cite{Shibauchi06a}) which is larger than the Pauli
paramagnetic limiting field \cite{Schuck06a}. This suggests that
at very low temperatures exotic superconducting phases might be
stable at high magnetic fields \cite{Schuck06a}. In the rest of
the $H-T$ phase diagram conventional vortex physics is expected.

The first of these works \cite{Hiroi06a} reports a low-field
re-entrant behavior of the temperature at which resistance becomes
negligible. The re-entrance is detected for vortices moving in
particular crystallographic directions. This suggests that the
phenomenon has its origin in a pinning mechanism arising from the
specific crystal structure of KOs$_2$O$_6$, indicating a
resemblance to the intrinsic pinning mechanism detected in
high-T$_{\rm c}$ superconductors \cite{Feinberg&Doyle}.

The second of these works \cite{Shibauchi07a} reports a drastic
change in the spatial distribution of vortices when cooling
through a temperature $T_{\rm p}$. Specific heat measurements
established that KOs$_2$O$_6$ presents a first-order phase
transition at an almost field-independent temperature $T_{\rm p}
\sim 8$\,K \cite{Hiroi06a,Bruhwiler06a}. This transition has been
associated with the freezing of the "rattling" phonon mode arising
from the vibration of the K ion within the oversized Os-O cage
\cite{Hiroi07a}. The low temperature ($T<T_{\rm p}$) vortex phase
is characterized by a reduced vortex line energy
\cite{Shibauchi07a}, implying a decrease in the vortex-vortex
interaction energy. This study therefore raises the question of a
structural transition in the KOs$_2$O$_6$ vortex matter occurring
at $T_{\rm p}$.

The third of these works \cite{Dubois07a} reports on Scanning
Tunnelling Microscopy imaging of vortices in the low temperature
phase. The observed structures present significant variations in
the intervortex distances. In particular, the spacing between some
vortices is roughly half the average vortex lattice parameter.
These findings are in agreement with a reduction of the vortex
interaction energy for the phase located at $T<T_{\rm p}$.

These works suggest that vortex matter in KOs$_{2}$O$_{6}$
presents unexpected properties for a low-T$_{\rm c}$ material.
Therefore, studies in RbOs$_{2}$O$_{6}$ and CsOs$_{2}$O$_{6}$ are
necessary to gain more insight into the properties of vortex
matter in the $\beta$-pyrochlore osmate family. This is
particularly relevant since Rb and Cs $\beta$-pyrochlores do not
present the first-order phase transition associated with a
dramatic change in the phonon spectra in the superconducting phase
\cite{Hiroi05b}.

In this work we study the vortex phase diagram of
RbOs$_{2}$O$_{6}$ single-crystals by means of bulk magnetization
measurements. Up until now, only one work reports on structural
and superconducting properties of RbOs$_{2}$O$_{6}$ single
crystals \cite{Rogacki08a} and the rest of the literature is
devoted to polycrystalline samples. The most important result
reported here is that RbOs$_{2}$O$_{6}$ presents a wide reversible
vortex region spanning down to a field $H_{\rm irr}(T) \sim 0.3
H_{\rm c2}(T)$ at low temperatures. This finding is in direct
contrast with results found in other low-T$_{\rm c}$
superconductors. We provide evidence that the wide reversible
region originates from the small critical current density in
RbOs$_{2}$O$_{6}$. The unusually wide reversible region and the
low critical current density observed in RbOs$_{2}$O$_{6}$ are
consistent with the available data for KOs$_{2}$O$_{6}$
\cite{Schuck06a}.

\section{Experimental and sample details}

The study presented here was carried out on two RbOs$_2$O$_6$
single crystals of the same batch that provided similar results.
The samples were grown in evacuated quartz ampoules following the
method described in Ref.\,\cite{Rogacki08a}. The crystals have a
prism-like shape with typical dimensions $0.1-0.2 \times 0.2
\times 0.2$\,mm$^3$ and weight $49.7 - 100.2 \pm 0.1$\,$\mu$g.

\begin{figure}
\begin{center}
\includegraphics[angle=-90,width=\textwidth]{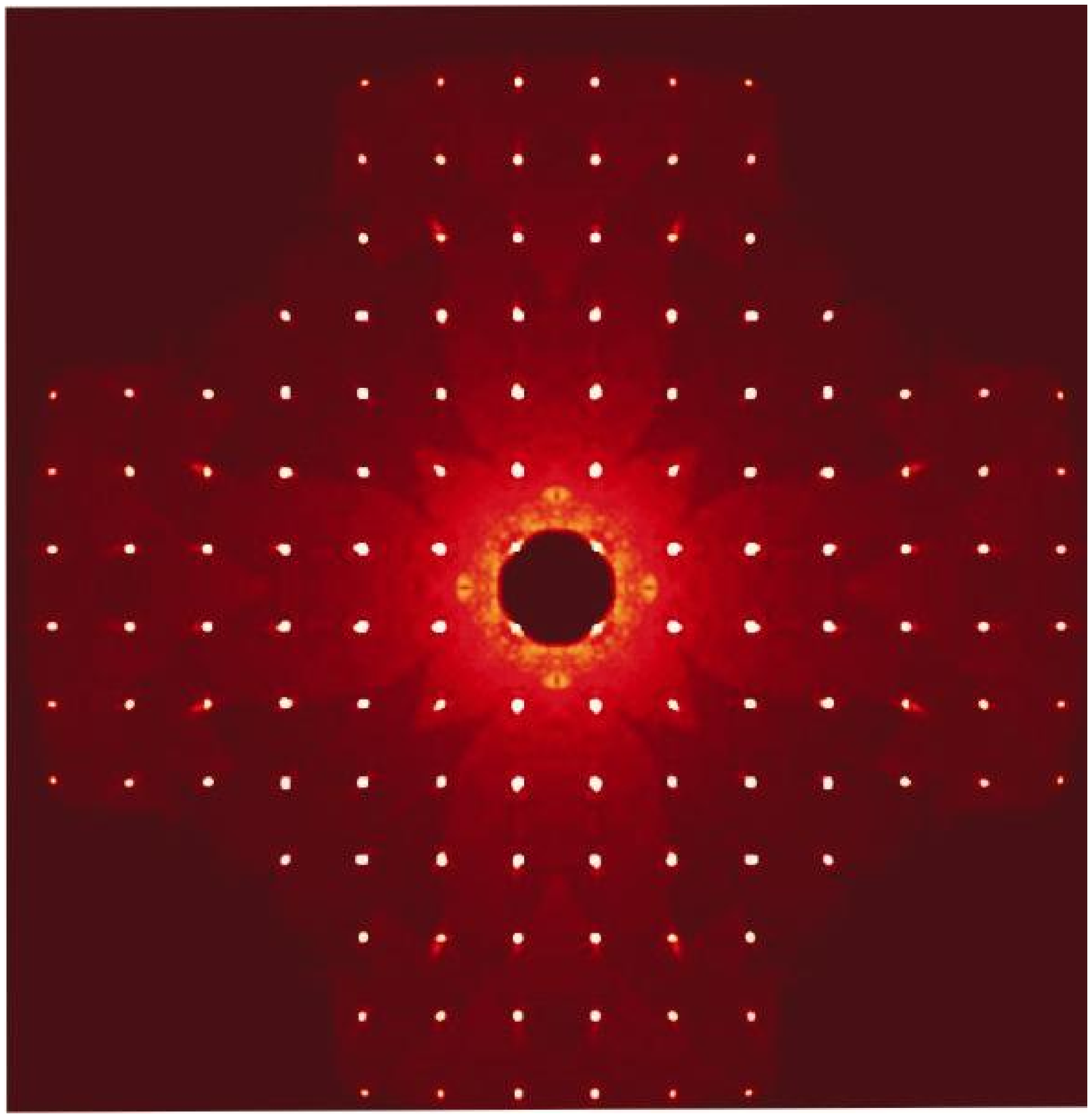}
\caption{Reconstructed $h1l$ reciprocal space section of a
RbOs$_2$O$_6$ sample measured at 295 K. \label{fig:RSMap}}
\end{center}
\end{figure}

The structural properties of the crystals were investigated on a
four-circle X-ray diffractometer (XCalibur PX of Oxford
Diffraction with an oscillation angle of 1$^{\circ}$ and Mo
K$\alpha$ radiation) equipped with a CCD area detector placed at
60\,mm from the sample. The data was refined on F$^2$ by employing
the program SHELXL-97 \cite{Sheldrick97a} and the results reveal a
$\beta$-pyrochlore cubic structure\cite{Yamaura06a, Galati07a}
(see Table\,\ref{table2}). The occupation of all elements remained
100\,\% during the structural refinement. No additional phases
(impurities, twins or intergrowth crystals) were detected by
examining the reconstructed reciprocal space section shown in
Fig.\,\ref{fig:RSMap}. The crystals present low mosaicity with an
average mosaic spread of 0.13(3)° (estimated analyzing every frame
by using XCalibur with the CrysAlis Software System
\cite{Oxford03a}). The observed reflections present a smaller
intensity than the calculated ones (extinction coefficient
$\epsilon = 64(7) \times 10^{-4}$, see Table\,\ref{table2}). This
can be explained by the very small misorientation of mosaic
blocks. The reconstructed reciprocal space section and refinement
results indicate that the single crystals used in this study are
of a high crystalline perfection.

\begin{table}[ttt]
\caption{Structure and refinement data for RbOs$_{2}$O$_{6}$ at
295,0(5)\,K.\label{table2}}
\begin{ruledtabular}
\begin{tabular}{cc}

Wavelength, \AA/radiation      &      0.71073/Mo K$\alpha$     \\
Crystal system, space group      &      cubic, Fd-3m (No 227)     \\
Unit cell dimensions, \AA      &      a =10.1214(8)     \\
Volume, \AA$^3$    &      1036.9(4)     \\
Z      &      8     \\
Absorption correction type      &      analytical     \\
Theta range for data collection, deg.      &      5.79 to 36.13     \\
Limiting indices      &      -13$\leqslant$h$\leqslant$12, -14$\leqslant$k$\leqslant$10, -13$\leqslant$l$\leqslant$13     \\
Reflections collected/unique      &      937/94, R$_{int}$ = 0.032     \\
Refinement method      &      Full-matrix least-squares on F$^2$     \\
Data /restraints/parameters      &      94/0/7     \\
Goodness-of-fit on F$^2$      &      1.310     \\
Final R indices [I $>$ 2sigma(I)]      &      R$_1$ = 0.0322, wR$_2$ = 0.0791     \\
R indices (all data)      &      R$_1$ = 0.0335, wR$_2$ = 0.0808     \\
Extinction coefficient      &      0.0064(7)     \\
$\Delta\rho_{max} , \Delta\rho_{min}$(e/\AA$^3$)      &      2.890 and -2.408     \\

\end{tabular}
\end{ruledtabular}
\end{table}

The superconducting magnetic properties of the samples were
characterized by zero-field-cooled (ZFC) and field-cooled (FC)
magnetization \textit{vs.} temperature  measurements, $M(T)$, and
magnetization \textit{vs.} magnetic field loops, $M(H)$.
Measurements for applied fields below 1.2\,T were performed in a
SQUID magnetometer; for larger applied fields a PPMS magnetometer
was used.

\section{Results and discussion}

\begin{figure}
\begin{center}
\includegraphics[angle=-90,width=\textwidth]{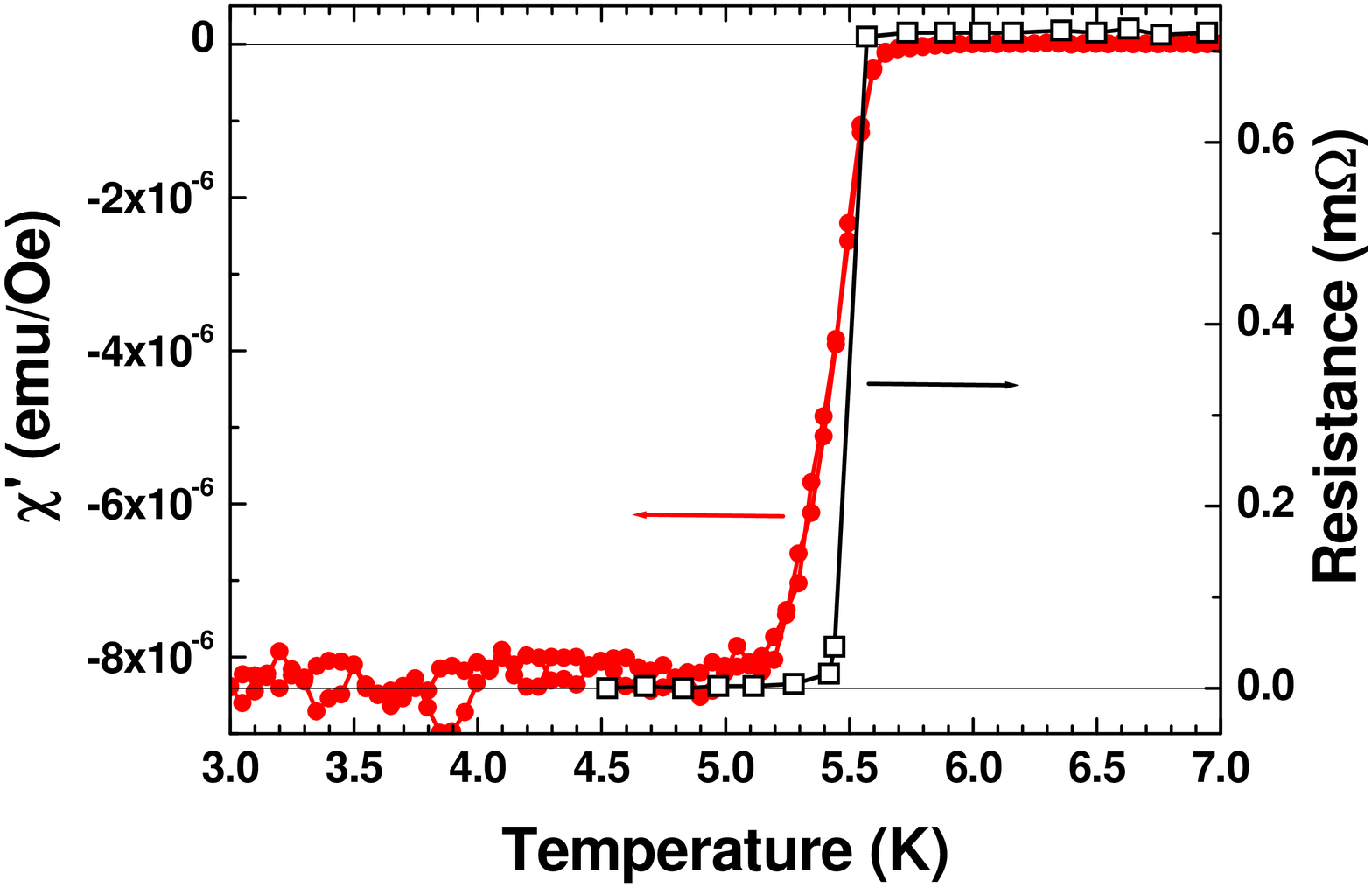}
\caption{(a) Superconducting transition of one of our
RbOs$_2$O$_6$ crystals at zero field as detected by resistivity
(open black symbols) and the real component of the AC
susceptibility (full red symbols). AC Susceptibility measurements
were performed with an applied field of 5\,Oe at 970\,Hz.
\label{fig:Tc}}
\end{center}
\end{figure}

The critical temperature of the samples was determined by
resistivity $R$, AC~susceptibility $\chi'$ and low-field
magnetization $M$ measurements (see Fig.\,\ref{fig:Tc}). T$_{\rm
c}$ is defined as the temperature at which $\partial
\chi'/\partial T$, $\partial M/\partial T$ and $\partial
R/\partial T$ present a peak. The transition width is estimated as
the full-width at half-maximum of these peaks. Critical
temperature values of T$_{\rm c}=(5.50 \pm 0.05)$\,K from
resistivity and $(5.45 \pm 0.03)$\,K from susceptibility and
magnetization were obtained. The transition width detected by
susceptibility or magnetization (0.3\,K) with an applied field of
11\,Oe is slightly broader than the one detected by resistivity
(0.2\,K) at zero field. A previous work \cite{Rogacki08a} reports
a higher T$_{\rm c}$ value for RbOs$_{2}$O$_{6}$ single-crystals,
but this property is known to be strongly dependent on the sample
disorder or small natural variations in stoichiometry.

\begin{figure}
\begin{center}
\includegraphics[angle=-90,width=\textwidth]{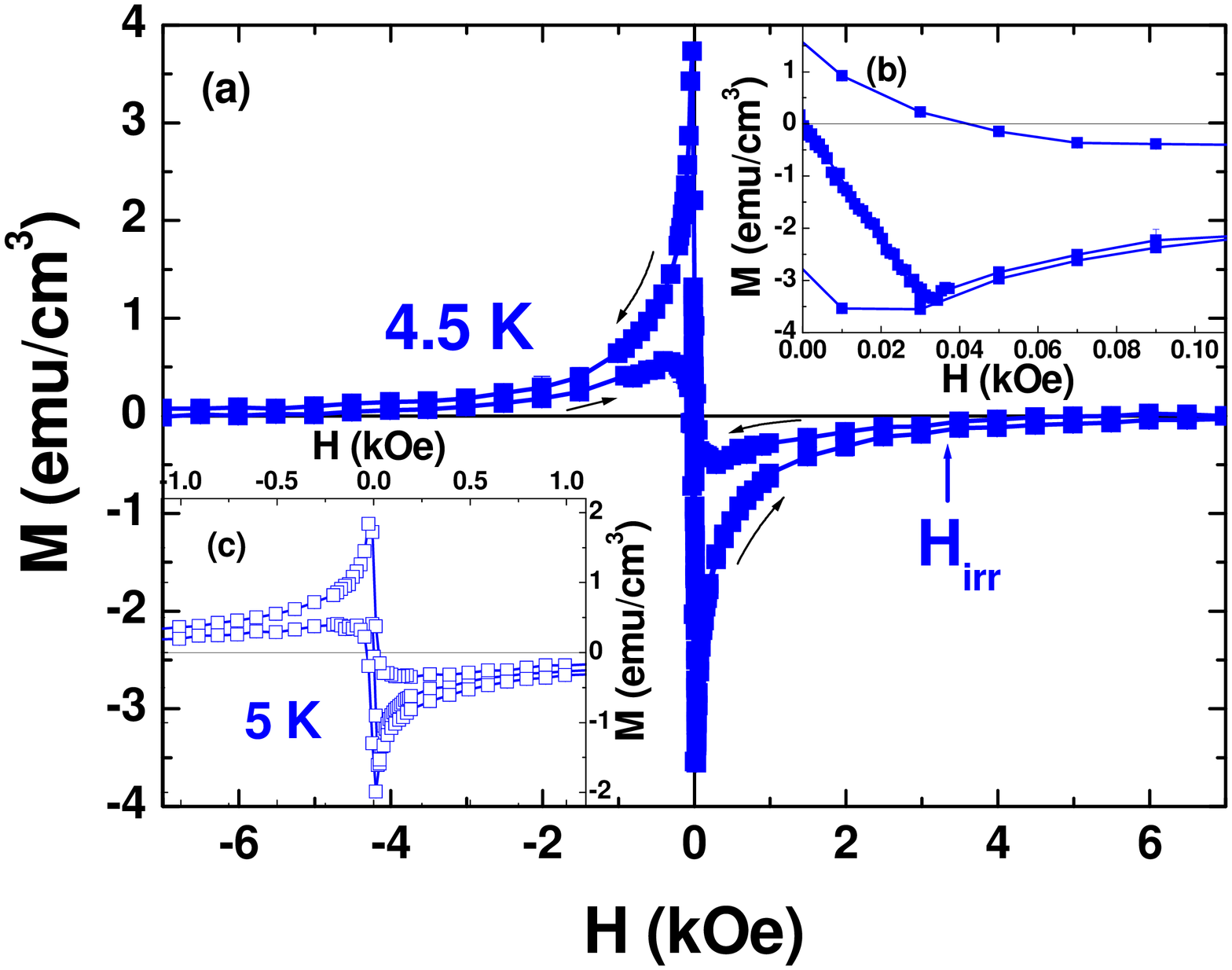}
\caption{Magnetization \textit{vs.} magnetic field curves for
RbOs$_2$O$_6$ single crystals. (a) Magnetization loop at 4.5\,K.
The irreversibility field determined from ZFC-FC M(T) measurements
is indicated. (b) Zoom of the magnetization in the low-field
region allowing a detailed observation of the Meissner branch. (c)
Magnetization loop of RbOs$_2$O$_6$ at 5\,K.\label{fig:Magloop}}
\end{center}
\end{figure}

The superconducting fraction of the sample was obtained from the
Meissner slope of the virgin branch of $M(H)$ loops such as the
one shown in Fig.\,\ref{fig:Magloop}b. After correcting for
demagnetization effects, we estimated our samples have a
superconducting fraction of $85 - 100$\,\%. Together with the fact
that the onset of the superconducting transition detected from
susceptibility measurements coincides with the temperature at
which resistance becomes negligible, this result indicates that
the samples undergo a bulk superconducting transition.

\begin{figure}
\begin{center}
\includegraphics[angle=-90,width=\textwidth]{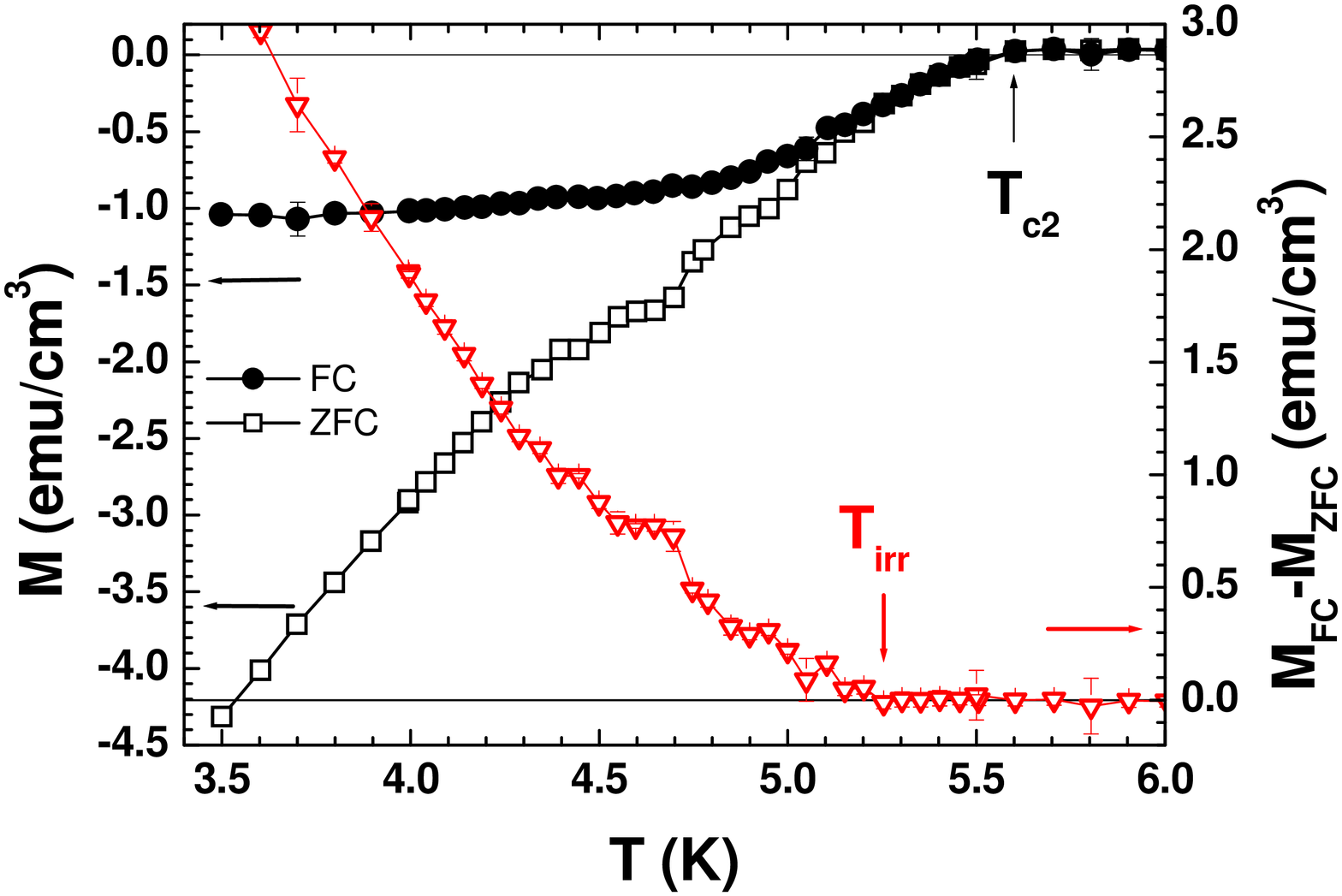}
\caption{Magnetization \textit{vs.} temperature curves of
RbOs$_2$O$_6$ following zero-field-cooling (ZFC) and field-cooling
(FC) processes at an applied field of 100\,Oe. The difference
between both branches is considered in order to estimate the onset
of the irreversible magnetic behavior at a temperature $T_{\rm
irr}(H)$. The upper critical temperature $T_{\rm c2}(H)$ is
estimated from the onset of screening. \label{fig:fczfc100G}}
\end{center}
\end{figure}

In order to determine the $H-T$ vortex phase diagram of
RbOs$_{2}$O$_{6}$ we obtained the upper critical field line
$H_{\rm c2}(T)$ and the irreversibility line $H_{\rm irr}(T)$ from
FC-ZFC $M(T)$ measurements. Typical $M(T)$ curves for an applied
field of 100\,Oe are shown in Fig.\,\ref{fig:fczfc100G}. For all
applied fields the value of the ZFC magnetization at low
temperatures gives a superconducting fraction consistent with the
one estimated from the Meissner slope. The temperature $T_{\rm
c2}(H)$ is determined from the onset of the superconducting
behavior in the ZFC and FC branches, as indicated in
Fig.\,\ref{fig:fczfc100G}. The temperature at which the vortex
magnetic response becomes irreversible on cooling, $T_{irr}(H)$,
is identified as the point at which both branches merge. We
obtained this temperature by plotting the difference $M_{\rm FC} -
M_{\rm ZFC}$, see Fig.\,\ref{fig:fczfc100G}.  For each applied
field, no difference in the values of T$_{\rm irr}$ was detected
when measuring at sweep rates of 25 and 5\,mK per minute.
Therefore, within this measurement-timescale the obtained values
of T$_{\rm irr}$ are not influenced by any dynamical effects.

\begin{figure}
\begin{center}
\includegraphics[angle=-90,width=\textwidth]{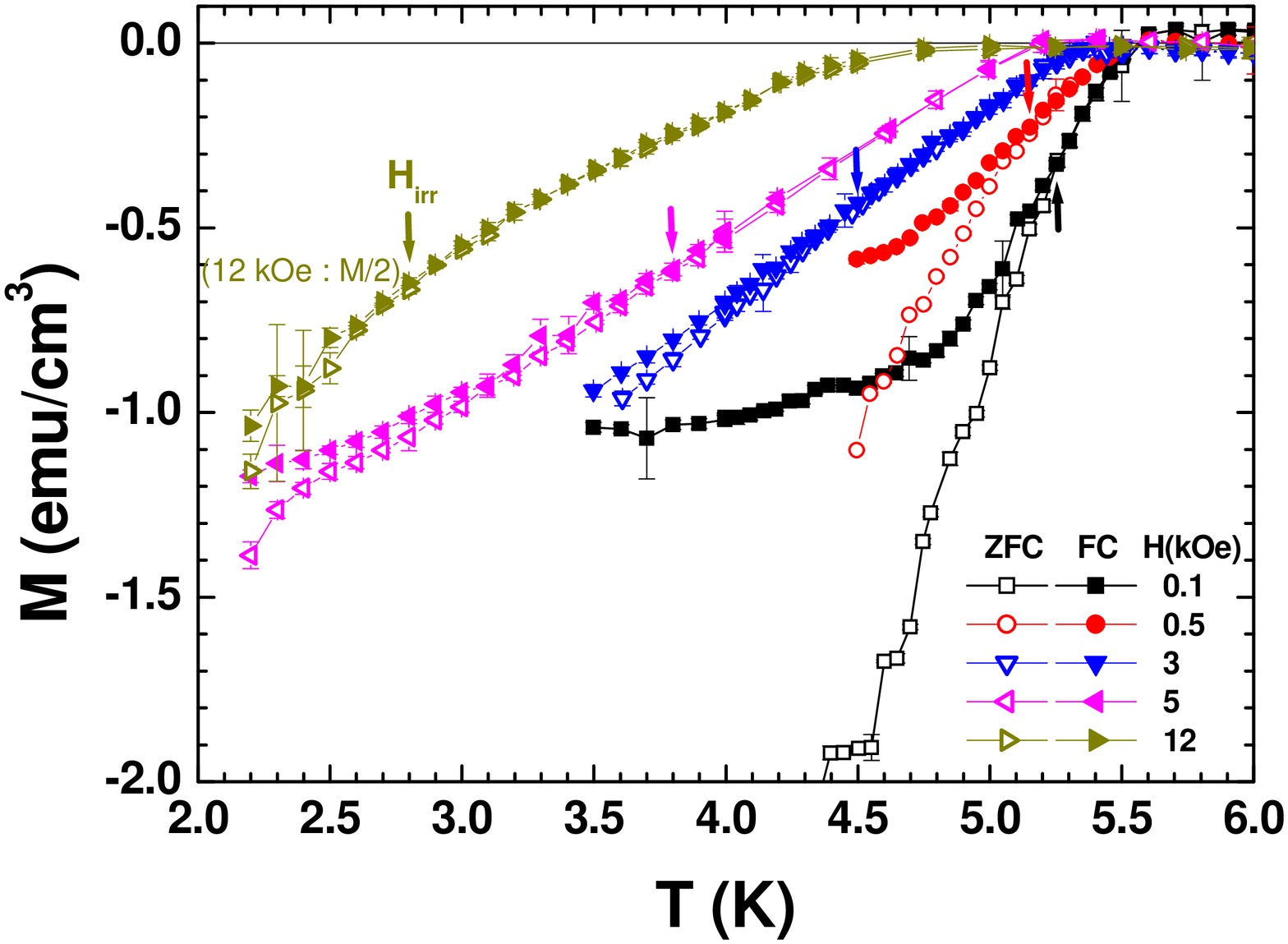}
\caption{Magnetization \textit{vs.} temperature curves of
RbOs$_2$O$_6$ following zero-field-cooling (open symbols) and
field-cooling (full symbols) processes for various applied fields.
 The irreversibility temperature at each field
$T_{irr}(H)$ is indicated with arrows. The measurements performed
at 12\,kOe are shown divided by a factor of two for clarity.
\label{fig:fczfcAll}}
\end{center}
\end{figure}

\begin{figure}
\begin{center}
\includegraphics[angle=-90,width=\textwidth]{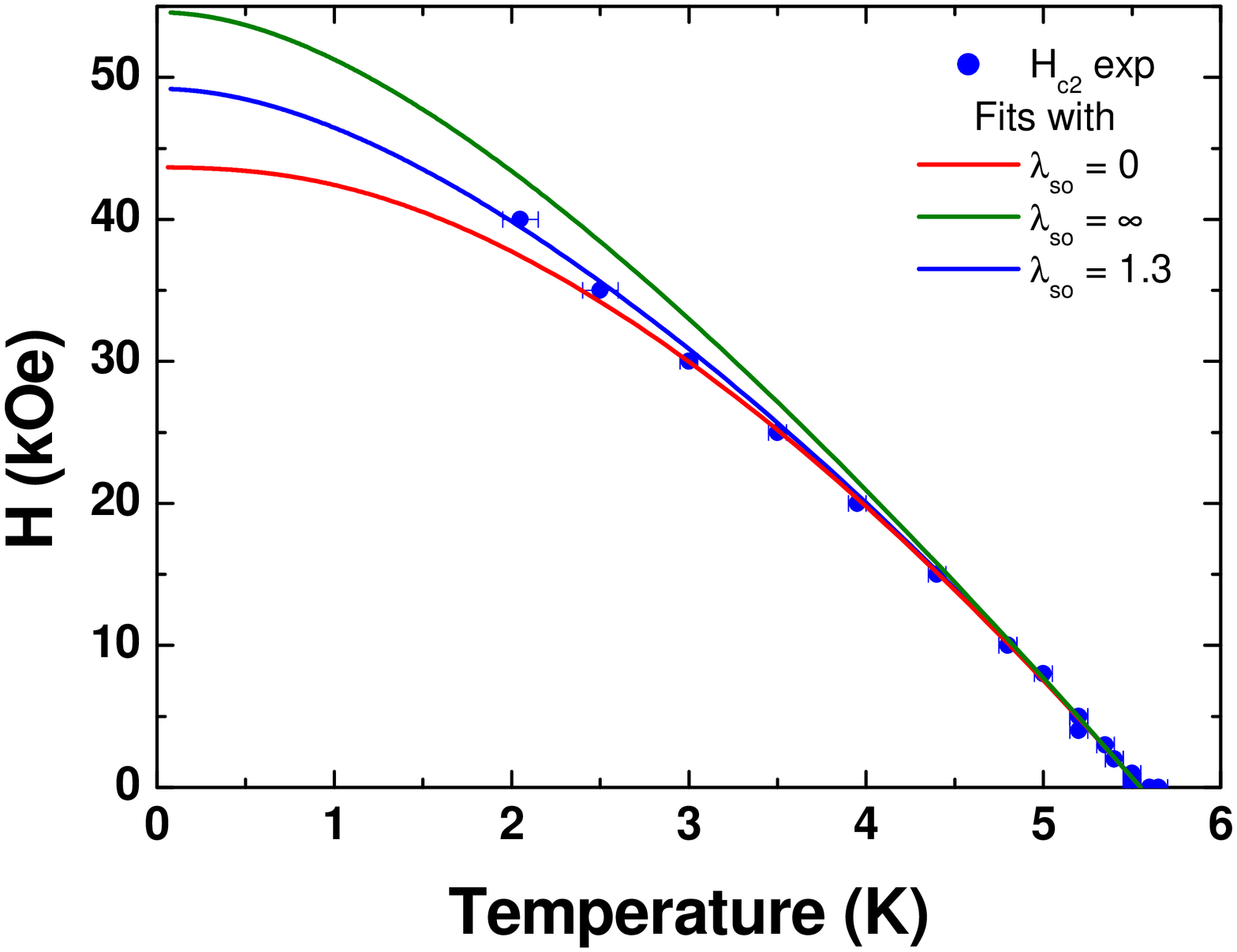}
\caption{Fits of H$_{c2}(T)$ with the WHH
model\,\cite{Helfand66a,Werthamer66a}. The experimental data is
shown with blue symbols. The red line corresponds to the zero
spin-orbit coupling case ($\lambda_{so}=0$) whereas the green line
is calculated using an infinite spin-orbit coupling
($\lambda_{so}=\infty$). The blue line is the best fit with a
finite spin-orbit coupling of $\lambda_{so}=1.3 \pm 0.2$.
\label{fig:Hc2Fits}}
\end{center}
\end{figure}

\begin{figure}
\begin{center}
\includegraphics[angle=-90,width=\textwidth]{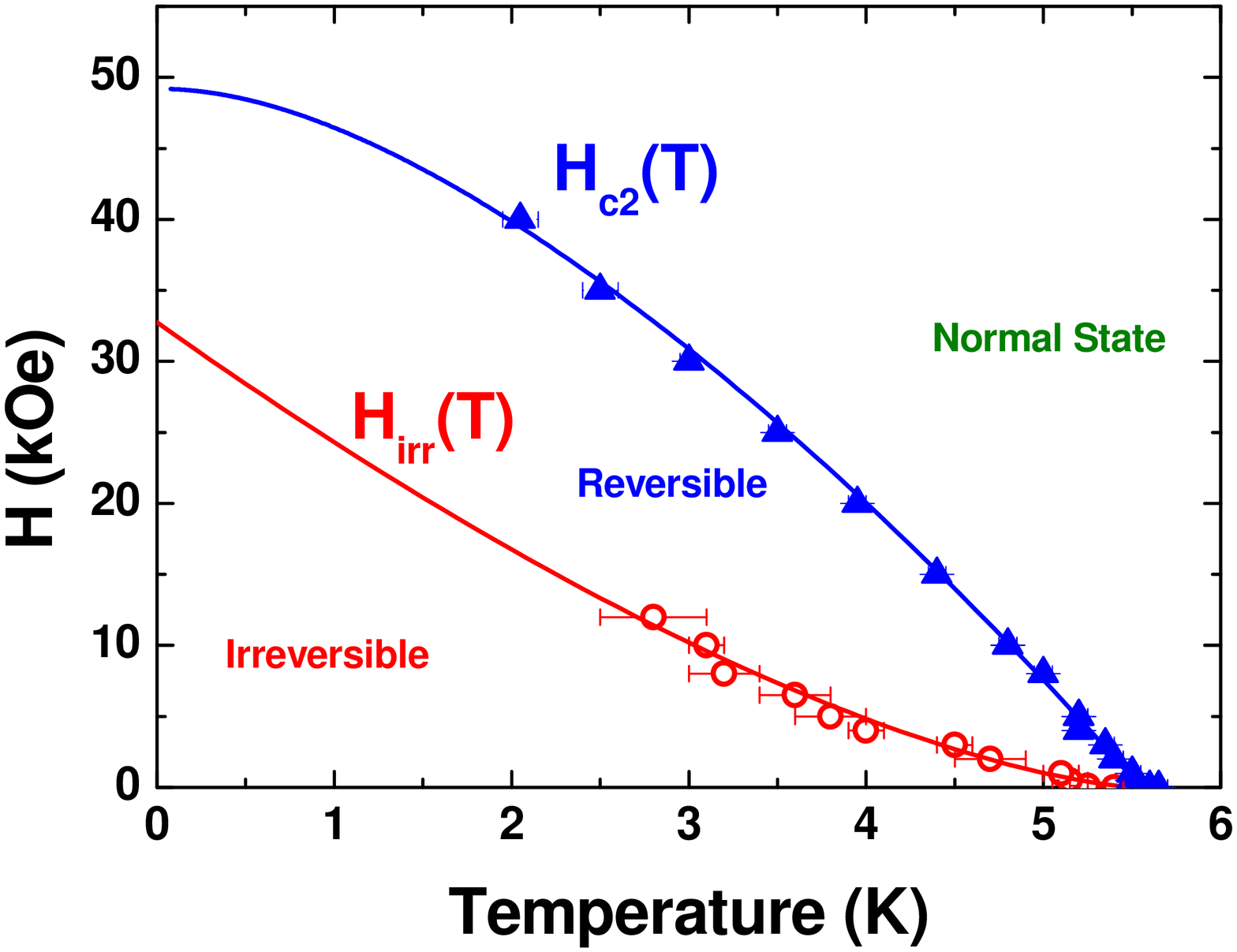}
\caption{Vortex phase diagram of RbOs$_2$O$_6$ : The upper
critical field line, $H_{\rm c2}(T)$, and the irreversibility
line, $H_{\rm IL}(T)$, are shown. The blue line is the best fit of
the data using the WHH model\,\cite{Helfand66a,Werthamer66a} which
yields $H_{\rm c2}(0) = 50$\,kOe. The red line is a fit of H$_{\rm
irr}$ with a temperature dependence proportional to $(1 - T/T_{\rm
c})^{1.5}$. \label{fig:vortexphase}}
\end{center}
\end{figure}

Examples of ZFC-FC $M(T)$ curves at various applied fields are
shown in Fig.\,\ref{fig:fczfcAll}. From the values of $T_{\rm
c2}(H)$  we obtained the upper critical field $H_{\rm c2}(T)$
indicated in Fig.\,\ref{fig:vortexphase}. In order to estimate the
zero-temperature upper critical field, $H_{\rm c2}(0)$, we fit
$H_{\rm c2}(T)$ with the Werthamer-Helfand-Hohenberg model (WHH)
\cite{Abrikosov61a,Helfand66a,Werthamer66a}. In the case of
RbOs$_{2}$O$_{6}$ the Pauli paramagnetic critical field
\cite{Clogston62a}  $H_{\rm P} \sim 18.4 T_{\rm c} = 101$\,kOe is
much larger than the $H_{\rm c2}(0)$ obtained by linearly
extrapolating $H_{\rm c2}(T)$ down to zero temperature ($\sim
60$\,kOe), indicating a strong spin-orbit coupling. This last
condition is fulfilled when $\alpha H/H_{\rm c2}(0) \ll
\lambda_{\rm so}$, where $\lambda_{\rm so}$ is the spin-orbit
scattering constant and $\alpha$ the Maki parameter
\cite{Maki64a}. In this limit, and in the absence of magnetic
impurities, the Abrikosov-Gor'kov upper critical field equation
reads (see Ref.\,\cite{Fischer72} for an overview)

\begin{equation}
\rm{ln}\left(\frac{1}{t}\right) = \Psi\left(\frac{1}{2} +
\frac{\rho_{\rm AG}(t)}{2t}\right) - \Psi\left(\frac{1}{2}\right)
\end{equation}

\noindent where $\Psi$ is the digamma function, $t=T/T_{\rm c}$,
and $h_{\rm c2}(t)=H_{\rm c2}(t)/H_{\rm c2}(0)$. The universal
pair-breaking function

\begin{equation}
\rho_{\rm AG}(t) = h_{\rm c2}(t) + \frac{\alpha^2 (h_{\rm
c2}(t))^2}{\lambda_{\rm so}}
\end{equation}

\noindent depends on $\lambda_{\rm so}$ and $\alpha$. The latter
is estimated from the slope of $H_{\rm c2}(T)$ at the vicinity of
T$_{\rm c}$ : $\alpha = -0.0528 \frac{dH_{\rm c2}(T)}{dT}|_{T_{\rm
c}} = 0.75$ (H$_{\rm c2}$ in kOe) \cite{Werthamer66a}, whereas
$\lambda_{\rm so}$ is a free parameter in the fitting procedure.
The best fit to the data with $\lambda_{\rm so} = 1.3 \pm 0.2$ is
presented in Fig.\,\ref{fig:Hc2Fits}.

The same figure shows that considering a finite spin-orbit
coupling is necessary to properly fit our experimental data. The
curves obtained in the extreme cases of zero ($\lambda_{\rm so} =
0$) and infinite ($\lambda_{\rm so} = \infty$) spin-orbit coupling
are shown with red and green lines. The latter curve was obtained
using equations (1) and (2), whereas the former was obtained
considering the general Abrikosov-Gor'kov equation in the
$\lambda_{\rm so} = 0$ limit \cite{Fischer72}. It is evident that
these two curves fail to properly fit the low-temperature data.
The best fit to the data with $\lambda_{\rm so} = 1.3 \pm 0.2$
therefore indicates that in RbOs$_2$O$_6$ the spin-orbit coupling
can be considered to be moderately strong.

The fitted $H_{\rm c2}(0)= 50$\,kOe is roughly half $H_{\rm P}$.
Within the Abrikosov-Gor'kov theory we obtain a value of $H_{\rm
c2}(0)$ slightly smaller than that reported in
Ref.\,\cite{Rogacki08a} for other single-crystals. However, in
that study the zero-temperature upper critical field was estimated
from a fit of $H_{\rm c2}(T)$ with an empirical power law
\cite{Rogacki08a}.  Our fit within the Abrikosov-Gor'kov theory
yields a coherence length $\xi(0)= \sqrt{\Phi_{0} / 2 \pi H_{\rm
c2}(0)} = 81$\,\AA. We therefore estimate a Ginzburg-Landau
parameter $\kappa = 31$  by considering the penetration depth
value obtained from muon-spin-rotation measurements in
polycrystalline samples \cite{Khasanov05a}.

The vortex phase diagram of Fig.\,\ref{fig:vortexphase} also shows
the locus of the irreversibility line, $H_{\rm irr}(T)$, obtained
from FC-ZFC magnetization \textit{vs.} temperature curves. The
most remarkable result shown in Fig.\,\ref{fig:vortexphase} is
that the reversible magnetic response of RbOs$_{2}$O$_{6}$ spans
an uncommonly wide region of the $H-T$ phase diagram. At low
temperatures, $2.5<T<4$\,K, $H_{\rm irr}(T)$ is of the order of
$0.3 H_{\rm c2}(T)$. Figure\,\ref{fig:vortexphase} also shows that
the irreversibility line can be well fitted with a sub-quadratic
power law, $H_{\rm irr} \propto (1 - T/T_{\rm c})^{1.5}$. These
findings are in contrast with expectations for low-T$_{\rm c}$
superconductors. For example, NbSe$_{2}$  has a reversible region
constrained to the vicinity of $H_{\rm c2}(T)$
\cite{Henderson96a,Angurel97a,Mohan07a}. Furthermore, the
irreversibility line in RbOs$_2$O$_6$ follows the same temperature
dependence as those of typical high-T$_{\rm c}$ superconductors
\cite{Tinkham}. Although there is no systematic study of the
temperature evolution of the irreversibility line in KOs$_2$O$_6$
crystals in the literature, one work reports that at 5\,K ($t \sim
0.5$) the vortex response becomes reversible at fields higher than
10\,kOe ($H_{\rm irr} \sim 0.1 H_{\rm c2}$) \cite{Schuck06a}.
Therefore, KOs$_2$O$_6$ seems to present an even wider reversible
vortex region than RbOs$_2$O$_6$.

An irreversible magnetic response in superconductors can have
three different origins: bulk pinning, Bean-Livingston surface
barriers \cite{Bean64b} and geometrical barriers
\cite{Zeldov94a,Indenbom94a}. In general, macroscopic
magnetization measurements are not able to ascertain which of the
three contributions is dominant when measuring an irreversible
magnetic response that sets in at $H_{\rm irr} (T)$. However, by
conveniently modifying the sample geometry the effect of
geometrical barriers can be affected. In the particular case of
prism-like samples it has been shown that the effect of
geometrical barriers in $H_{\rm irr}(T)$ is negligible
\cite{Majer95a}. The Bean-Livingston surface barrier only produces
a significant irreversible behavior in the case of extremely
smooth surfaces \cite{DeGennes}. In real samples with sharp
corners and irregular edges, the effect of this barrier is of
lesser importance. Therefore, as the crystals studied in this work
are prism-like, $H_{\rm irr}(T)$ can be considered  as the field
at which point pinning sets in while cooling, i.e. the depinning
line. Strictly speaking, the effect of pinning may become relevant
at slightly lower fields than $H_{\rm irr}(T)$.

The depinning line is determined by the competition between
thermal fluctuations and pinning generated by quenched disorder
naturally present in the samples \cite{Blatter94a}. The magnitude
of quenched disorder is typically measured by the dimensionless
critical current-density ratio, $J_{\rm c}(T,H)/J_{\rm 0}(T)$,
with $J_{\rm 0}(T)= 4c \Phi_{\rm 0}/ 12 \sqrt{3} \pi
\lambda^{2}(T) \xi(T)$ the depairing current density
\cite{Blatter94a}. For low-T$_{\rm c}$ superconductors this
parameter is typically of the order of $10^{-2}-10^{-1}$ whereas
in high-T$_{\rm c}$ materials the pinning strength is weaker since
$J_{\rm c}/J_{\rm 0} \sim 10^{-5}-10^{-2}$ at low temperatures and
fields. The relevance of thermal fluctuations increases with the
Ginzburg number of the material, $G_{\rm i}= 0.5(k_{\rm B} T_{\rm
c} \gamma \kappa^2/H_{\rm c2}(0)^{2}\xi(0)^{3})^{2}$, proportional
to the electronic anisotropy $\gamma = \sqrt{m_{\rm c}/m_{\rm ab}}
\geq 1$ \cite{Blatter94a}. Typically, $G_{\rm i} \sim
10^{-4}-10^{-8}$ for low-T$_{\rm c}$ and $\sim 10-10^{-2}$ for
high-T$_{\rm c}$ superconductors \cite{Blatter94a}. Therefore,
both a small critical-current density ratio and a large Ginzburg
number can conspire to produce a wide reversible vortex region.

The Ginzburg number $G_{\rm i} = 6 \times 10^{-7}$ obtained for
RbOs$_2$O$_6$ from the $H_{\rm c2}(0)$ estimated in this work is
within the range of values typically found for low-T$_{\rm c}$
materials. To obtain this value we assumed a negligible electronic
anisotropy ($\gamma = 1$) based on the reported isotropic carrier
mass for KOs$_2$O$_6$\cite{Hiroi07a} and the absence of similar
data for RbOs$_2$O$_6$. The value of $G_{\rm i}$ for RbOs$_2$O$_6$
indicates that thermal fluctuations are conventional and cannot
account for the wide reversible vortex region. The Ginzburg number
of KOs$_2$O$_6$ is one order of magnitude larger than that of
RbOs$_2$O$_6$. This is a consequence of $H_{\rm c2}(0)$ ($\xi(0)$)
being larger (smaller) in KOs$_2$O$_6$ (see Table\,\ref{table1}
for numerical details). However, this larger $G_{\rm i}$ cannot
account for the greater extent of the reversible vortex region in
KOs$_2$O$_6$: for example $H_{\rm irr}(t=0.5)/H_{\rm c2}(t=0.5)
\sim 0.1$  and  $ \sim 0.35$, for KOs$_2$O$_6$ and RbOs$_2$O$_6$
respectively.

For illustrative purposes, it is very interesting to compare the
case of RbOs$_2$O$_6$ with that of NbSe$_2$. Both compounds have
similar T$_{\rm c}$, $\lambda$ and $\xi$, but NbSe$_2$ is more
anisotropic with $\gamma = 3.3$, resulting in a Ginzburg number
one order of magnitude larger than that of RbOs$_2$O$_6$. However,
the reversible vortex region in NbSe$_2$ is constrained to 0.1\,K
below $H_{\rm c2}(T)$ \cite{Mohan07a} whereas in the case of
RbOs$_2$O$_6$ it is much wider.

\begin{figure}
\begin{center}
\includegraphics[angle=-90,width=\textwidth]{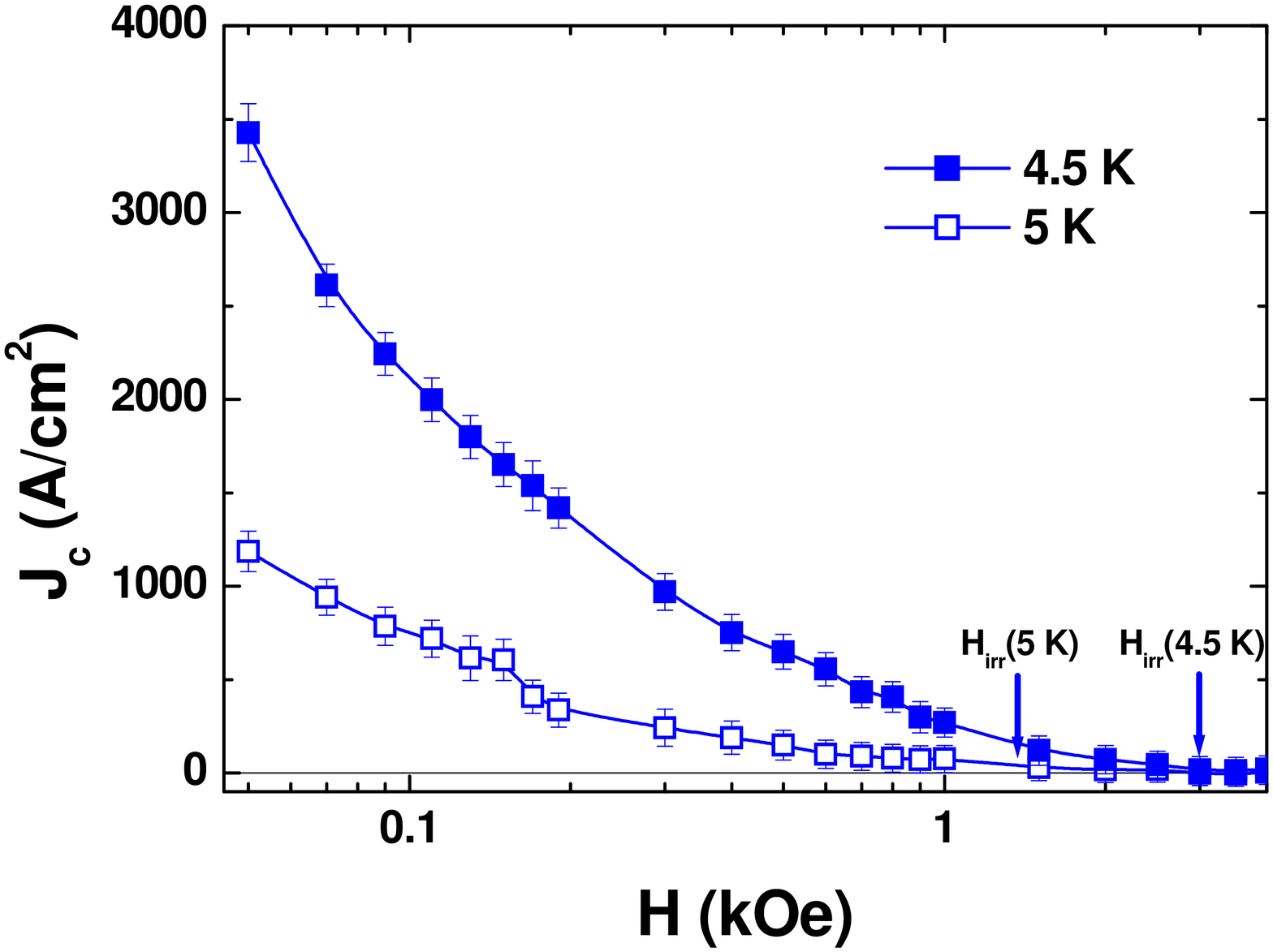}
\caption{Critical current density as a function of magnetic field
for RbOs$_2$O$_6$ at 4.5\,K (full squares) and 5\,K (open
squares). The irreversibility fields determined from FC-ZFC M(T)
measurements are indicated. \label{fig:Jc}}
\end{center}
\end{figure}

As a consequence, the wide reversible vortex region of
RbOs$_2$O$_6$ has to be caused by a low critical current density.
We estimated the critical current density, $J_{\rm c} (T,H)$,
assuming that the effect of surface and geometrical barriers is
negligible. In this case, within the Bean model \cite{Bean64a} the
critical current density can be estimated from $M(H)$ loops as
$J_{\rm c}(T,H) \sim (c/f) \Delta M(T,H)$. Here $\Delta M(T,H)$ is
the separation between the two branches of the magnetization loop
at a field $H$, $c$ the speed of light, and $f = (a/2)(1-a/3b)$,
where $a$ and $b$ are the dimensions in the plane perpendicular to
the applied magnetic field. Figure\,\ref{fig:Jc} shows the
$J_{c}(H)$ curves for temperatures of 4.5 and 5\,K. As expected,
the critical current density decreases with magnetic field and
temperature and consistently becomes negligible at the
irreversibility field determined from FC-ZFC magnetization
measurements.

\medskip

\medskip

\begin{table}[ttt]
\caption{Measured and derived superconducting parameters for the
$\beta$-pyrochlores RbOs$_{2}$O$_{6}$ and KOs$_{2}$O$_{6}$ and the
low and high-T$_{\rm c}$ superconductors NbSe$_2$ and
optimally-doped
Bi$_{2}$Sr$_{2}$Ca$_{2}$Cu$_{3}$O$_{10}$.\label{table1}}
\begin{ruledtabular}
\begin{tabular}{ccccc}

Parameter & RbOs$_{2}$O$_{6}$ & KOs$_{2}$O$_{6}$ & NbSe$_{2}$ & Bi$_{2}$Sr$_{2}$Ca$_{2}$Cu$_{3}$O$_{10}$\\
\hline

$T_{\rm c}\,[K]$ & 5.5\,$^{a}$ & 9.6\,$^{d}$ & 7.2\,$^{h,i}$ & 110.5\,$^{k}$\\
$\xi(0)$\,[\AA] & 81\,$^{a}$ & 31-37\,$^{d}$ & 77\,$^{h}$ & $\sim 10\,^{k}$\\
$\lambda(0)$\,[\AA] & 2500\,$^{b}$ & 2500-2700\,$^{d,e}$ & 2000\,$^{i}$ & 500\,$^{k}$\\
$\kappa=\frac{\lambda(0)}{\xi(0)}$ & 31 & 70-87 & 26 & $\sim 50$\\
$\gamma$ & - \,$^{c}$ & $\sim 1$\,$^{f}$& 3.3\,$^{h}$ &  27\,$^{k}$\\
$H_{\rm c2}(0)$\,[kOe] & 50\,$^{a}$& 340\,$^{d}$& 55\,$^{h}$& $\sim 3000$\\
$G_{\rm i}=0.5(\frac{k_{\rm B} T_{\rm c} \gamma \kappa^2}{H_{\rm
c2}(0)^{2}\xi(0)^{3}})^{2}$ & 5\,10$^{-7}$\,$^{c}$& 5\,10$^{-6}$ & 5\,10$^{-6}$& 2\,10$^{-2}$\\
$J_{\rm c}(t,h(T))/J_{0}(t)$ & 5\,10$^{-5}$\,$^{a}$& 5\,10$^{-6}$\,$^{g}$ & 3\,10$^{-1}$\,$^{j}$ & 1\,10$^{-5}\,^{k}$\\
for $h(T)=0.02$ and $t=$ & 0.81& 0.52 & 0.59 & 0.16\\

\end{tabular}
\end{ruledtabular}
\end{table}

$^{a}$ This work, single crystals.

$^{b}$ Ref.\,\cite{Khasanov05a}, polycrystals.

$^{c}$ No data available in the literature. In order to calculate
$G_{\rm i}$ we assumed $\gamma=1$ (see text).

$^{d}$ Ref.\,\cite{Bruhwiler06a}, single crystals.

$^{e}$ Ref.\,\cite{Koda05a}, polycrystals.

$^{f}$ Ref.\,\cite{Hiroi07a}, single crystals.

$^{g}$ Critical current calculated by us from data of
Ref.\,\cite{Schuck06a}.

$^{h}$ Ref.\,\cite{Trey73a}.

$^{i}$ Ref.\,\cite{Takita&Le}.

$^{j}$ Ref.\,\cite{Angurel97a}.

$^{k}$ Ref.\,\cite{Piriou08a}.

According to the results in Fig.\,\ref{fig:Jc}, for RbOs$_2$O$_6$
the critical-current density ratio $J_{\rm c}/J_{0} \sim 5 \times
10^{-5}$ for reduced temperature $t=4.5/T_{\rm c}=0.81$ and field
$h(T)=H/H_{\rm c2}(T)=0.02$. To estimate this ratio we have
calculated $J_{0}(t)$ considering the two-fluid model expression
for $\lambda(T)$ and $\xi(T)$ \cite{Blatter94a}. A similar
critical current density ratio is obtained for a reduced
temperature $t=0.91$. These values of $J_{\rm c}/J_{0}$ for
RbOs$_2$O$_6$ are smaller than values typically measured for other
low-T$_{\rm c}$ materials. For example, they are four orders of
magnitude smaller than that of NbSe$_2$ at the same reduced field
\cite{Angurel97a}. Strikingly, the value of $J_{\rm c}/J_{0}$ is
comparable to that of high-T$_{\rm c}$ compounds: for example it
is of a similar order of magnitude to that of
Bi$_{2}$Sr$_{2}$Ca$_{2}$Cu$_{3}$O$_{10}$ at low fields and
temperatures (see Table\,\ref{table1}). Therefore, in
RbOs$_2$O$_6$ quenched disorder has an importance as small as in
the case of high-T$_{\rm c}$'s. In spite of T$_{\rm c}$ being much
smaller, the low value of $J_{\rm c}(T,H)/J_{0}(T)$ is at the root
of the unusually wide reversible vortex region detected in
RbOs$_2$O$_6$.

A low magnitude of the critical-current density ratio might be
generic to the $\beta$-pyrochlore family. In the case of
KOs$_2$O$_6$, although no data on $J_{\rm c}(T,H)$ is available in
the literature, we have considered the $M(H)$ data of
Ref.\,\cite{Schuck06a} in order to estimate its critical current
density at 5\,K and $h(T)=0.02$. As shown in Table\,\ref{table1},
KOs$_2$O$_6$ has a $J_{\rm c}(T,H)/J_{0}(T)$ one order of
magnitude smaller than that of RbOs$_2$O$_6$. The decreased
relevance of quenched disorder would explain the suspected wider
reversible vortex region of KOs$_2$O$_6$ \cite{Schuck06a}.

\section{Conclusions}

In conclusion, we present the first data on the vortex matter
phase diagram in RbOs$_2$O$_6$ single crystals \cite{Rogacki08a}.
We found that this compound presents a reversible vortex region
that is unexpectedly wide for a low-T$_{\rm c}$ material. This
finding might be generic to the $\beta$-pyrochlore osmate
superconductors.

We found that this phenomenon originates from weak bulk pinning
since the relevance of thermal fluctuations seems to be limited.
The structural characterization results presented here suggest
that the crystal defect density is very low in RbOs$_2$O$_6$. This
can explain the weak pinning magnitude. Surprisingly, the Rb and K
members of the $\beta$-pyrochlore family present a
critical-current density ratio comparable to that of high-T$_{\rm
c}$ superconductors. Furthermore, resistivity measurements in
KOs$_2$O$_6$ single crystals suggest an intrinsic pinning
mechanism \cite{Hiroi06a}, a feature that is typically observed in
high-T$_{\rm c}$'s \cite{Feinberg&Doyle}. The evidence presented
here therefore indicates that the negligible importance of bulk
pinning produces a wide reversible region.

The authors acknowledge M. Decroux, F. de la Cruz, A. A.
Petrovi\'c and G. Santi for useful discussions and A. Piriou and
R. Lortz for assistance in the SQUID measurements. This work was
supported by the MaNEP National Center of Competence in Research
of the Swiss National Science Foundation.

\end{document}